\documentstyle[fleqn,aip2col,epsf]{article}
\let\addressmark\relax

\newcommand{\lsim}{\mathrel{\raisebox{-.6ex}{$\stackrel{\textstyle<}{\sim}$}}}
\newcommand{\gsim}{\mathrel{\raisebox{-.6ex}{$\stackrel{\textstyle>}{\sim}$}}}

\begin{document}
\pagestyle{plain}
\thispagestyle{empty}

\title{\vglue-.6in
\font\fortssbx=cmssbx10 scaled \magstep1
\hbox to \hsize{
\hbox{\fortssbx University of Wisconsin - Madison}
\hfill$\vcenter{\normalsize\hbox{\bf MADPH-97-1012}
                \hbox{August 1997}
                \hbox{\hfil}}$ }\smallskip
Physics at Muon Colliders\footnotemark}
\author{\unskip\smallskip
V. Barger\address{Physics Department, University of Wisconsin, Madison, WI 53706, USA}}

\begin{abstract}
The potential of muon colliders to address fundamental physics issues is explored, with emphasis on understanding the nature of electroweak symmetry breaking. The $s$-channel production of Higgs bosons, unique to a muon collider, along with precision measurements of $W^+W^-$, $t\bar t$, and $Zh$ thresholds would determine the properties of Higgs bosons and test electroweak radiative corrections. At the high energy frontier, a 4~TeV muon collider is ideally suited to study either the production of supersymmetric scalar particles or a strongly interacting $WW$ sector.
\end{abstract}

\maketitle

\renewcommand{\thefootnote}{\fnsymbol{footnote}}
\footnotetext{Talk present at the {\it FCP\,97 Workshop on Fundamental Particles and Interaction}, Vanderbilt University, May 1997.}

\section{Introduction}

Why should we be interested in muon colliders? The reasons derive from the large mass of the muon compared to the electron. Far less radiation from muons allows excellent energy resolution ($\sim$few MeV) and fine energy calibration ($\Delta E/E \sim 1$~ppm). These features, and the enhanced coupling of the Higgs boson to muons, make possible $s$-channel Higgs resonance studies. Moreover, multi-TeV muon colliders with high luminosity are feasible for the study of high threshold phenomena. 
 But muon colliders pose a challenging technology because of the short muon lifetime ($\tau_\mu \sim 2.2\times10^{-6}{\rm\,s}$, $c\tau_\mu\sim 660$~m), $\mu$-decay and rescattering backgrounds in the detector, and possible radiation hazards at high energies from neutrinos emitted in muon decays. Nonetheless, the physics merits of muon colliders fully warrant research and development towards their realization. The recent accelerator developments related to the design of muon colliders are summarized in the proceedings of a series of workshops\cite{workshops}. This report is devoted to a discussion of the physics that can be done at muon colliders, which has also been the subject of intense investigations\cite{physrep}.

Several muon collider designs are under consideration. The First Muon Collider (FMC) is presently envisaged to operate at a center-of-mass energy in the range from $M_Z$ to above the $t\bar t$ threshold with corresponding average luminosities of $10^{32}$ to $10^{33}\rm\,cm^{-2}\,s^{-1}$ (1 to 10~fb$^{-1}$/yr). The FMC would be a first step towards a very high energy collider, the Next Muon Collider, with c.m.\ energy of 3 to 4~TeV and luminosity $10^{35}\rm\,cm^{-2}\,s^{-1}$ (1000~fb$^{-1}$/yr).

Muon colliders would enable us to resolve the major physics issue of our time: how is the electroweak symmetry broken? There are two scenarios for the symmetry breaking. A weakly broken scenario is based on Higgs bosons, and the most popular version invokes a low energy supersymmetry (SUSY). A strongly broken electroweak  scenario (SEWS) most likely would involve new resonances at the TeV scale of a still unknown dynamics. Muon colliders offer unique physics opportunities such as $s$-channel Higgs production, precision threshold measurements, and high energy and luminosity for SEWS or SUSY studies of TeV mass scale particles.

\section{$s$-channel Higgs Production}

Higgs bosons can be produced in the $s$-channel with interesting rates at a muon collider, which is an unique highly advantageous prospect\cite{saus,prl}. The resonance cross section is
\begin{equation}
\sigma_h\left(\sqrt{\hat s}\right) = { 4\pi\Gamma(h\to\mu\mu)\, \Gamma(h\to X) \over \left(\hat s - m_h^2\right)^2 + m_h^2 \left[ \Gamma_h^{\rm tot} \right]^2 }\,,
\end{equation}
where $\hat s$ is the c.m.\ energy squared, $\Gamma_h^{\rm tot}$ is the total width, and $X$ denotes the final state. The Higgs coupling to fermions is proportional to the fermion mass so the corresponding $s$-channel process is highly suppressed at $e^+e^-$ colliders. The cross section must be convoluted with the energy resolution, approximated by a Gaussian distribution of width $\sigma_{\sqrt s}$,
\begin{eqnarray}
\bar\sigma_h\left(\sqrt s\right) = \int \sigma_h\left( \sqrt{\hat s} \right)
 {\exp  \left[- \displaystyle  { \left( \sqrt{\hat s} - \sqrt s \right)^2 \over \left(2\sigma_{\sqrt s}^2 \right)} \right] d\sqrt s\over\sqrt{2\pi} \,\sigma_{\sqrt s} }\,.
\end{eqnarray}
The root mean square spread $\sigma_{\sqrt s}$ in c.m.\ energy is given in terms of the beam resolution $R$ by
\begin{equation}
\sigma_{\sqrt s} = (7{\rm\ MeV}) \left(R\over 0.01\%\right) 
\left(\sqrt s\over 100\rm\ GeV\right) \,,
\end{equation}
where a resolution down to $R=0.003\%$ may be realized at the FMC. In comparison, values of $R>1\%$ are expected at a linear $e^+e^-$ collider. To study a Higgs resonance one wants to be able to tune the machine energy to $\sqrt s = m_h$. For this purpose the monochromaticity of the beam energy is vital.

When the resolution is much larger than the Higgs width, $\sigma_{\sqrt s} \gg \Gamma_h^{\rm tot}$, the effective $s$-channel cross section is
\begin{equation}
\bar\sigma_h = {2\pi^2 {\rm BF} (h\to\mu\mu) \, {\rm BF}(h\to X)\over \sqrt{2\pi} \, m_h^2} \cdot {\Gamma_h^{\rm tot}\over \sigma_{\sqrt s} } \,.
\end{equation}
In the other extreme of resolution much smaller than the width, $\sigma_{\sqrt s} \ll \Gamma_h^{\rm tot}$, the effective cross section is
\begin{equation}
\bar\sigma_h = {4\pi\Gamma(h\to\mu\mu) \, {\rm BF}(h\to X)\over m_h^2} \cdot
{1\over\Gamma_h^{\rm tot}} \,.
\end{equation}

Figure~1 illustrates the SM Higgs cross section for several choices of the machine resolution. The resolution requirements for the machine depend on the Higgs width. Figure~2 gives both SM and SUSY Higgs width predictions versus the Higgs mass. A SM Higgs of mass $m_h\sim100$~GeV has a width of a few MeV. The width of the lightest SUSY Higgs may be comparable to that of the SM Higgs (if $\tan\beta\sim1.8$) or much larger ($\Gamma_h\sim0.5$~GeV for $\tan\beta\sim50$~GeV).  Figure~3 shows light Higgs resonance profiles versus the c.m.\ energy $\sqrt s$. With a resolution $\sigma_{\sqrt s}$ of order $\Gamma_h$ the Breit-Wigner line shape can be measured and $\Gamma_h$ determined. For the moment we must plan for a resolution $R\lsim0.01\%$ in order to be sensitive to $\Gamma_h$ of a few MeV.

\begin{figure}[h]
\centering
\leavevmode
\epsfxsize=3in\epsffile{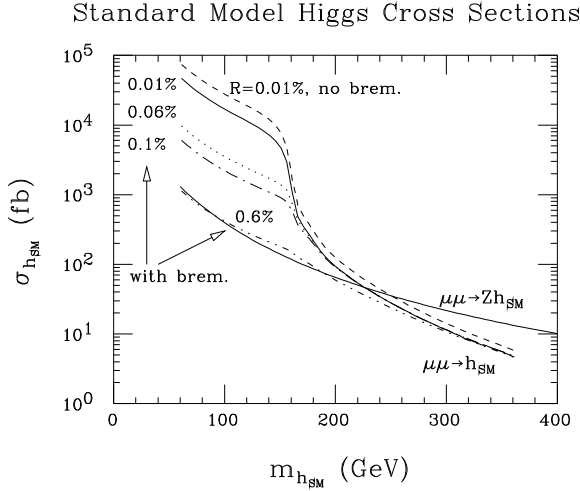}

\caption[]{The $s$-channel cross section for $\mu^+\mu^-\to h_{\rm SM}$ for several choices of the beam resolution $R$. Also shown is the $\mu^+\mu^-\to Zh_{\rm SM}$ cross section at $\sqrt s = M_Z + \sqrt2 m_{h_{\rm SM}}$. From Ref.~\cite{physrep}.}
\end{figure}

\begin{figure}[t]
\centering
\leavevmode
\epsfxsize=3in\epsffile{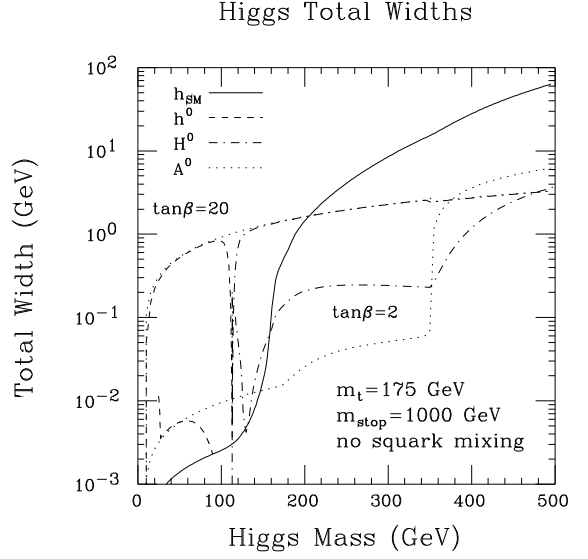}

\caption[]{Total width of the SM and MSSM Higgs bosons with $\tan\beta=2$ and 20. From Ref.~\cite{physrep}.}
\end{figure}

\begin{figure}[b]
\centering
\leavevmode
\epsfxsize=3in\epsffile{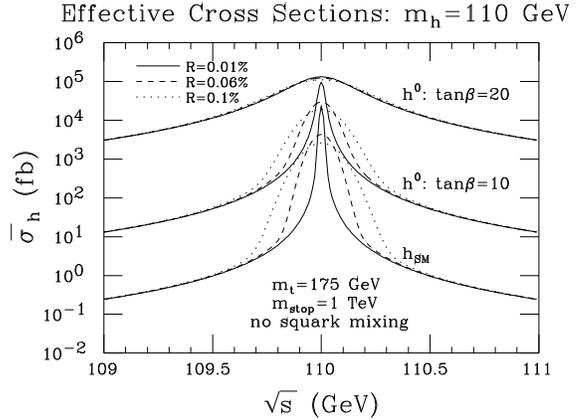}

\caption[]{Effective $s$-channel higgs cross section $\bar\sigma_h$ obtained by convoluting the Breit-Wigner resonance formula with a Gaussian distribution for resolution $R$. From Ref.~\cite{physrep}.}
\end{figure}

The prospects for observing the SM Higgs are evaluated in Fig.~4. The first two panels give the signal and background for a resolution $R=0.01\%$. The third panel gives the necessary luminosity for a $5\sigma$ detection in the dominant $b\bar b$ final state. The luminosity requirements are very reasonable, except for the $Z$-boson peak region. 

\begin{figure}[t]
\centering
\leavevmode
\epsfxsize=3in\epsffile{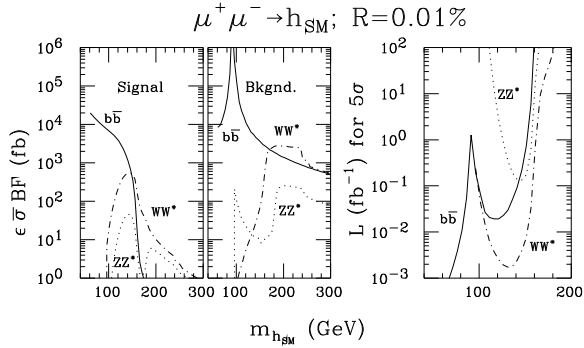}

\caption[]{The SM Higgs cross sections and backgrounds in $b\bar b,\ WW^*$ and $ZZ^*$. Also shown is the luminosity needed for a 5~standard deviation detection in $b\bar b$. From Ref.~\cite{physrep}.}
\end{figure}

\looseness=-1
From a rough scan for the $s$-channel $h^0$ signal the mass can be determined to an accuracy $\Delta m_h \sim\sigma_{\sqrt s}$. If $S/\sqrt B\gsim3$ is required for detection or rejection of a Higgs signal and a resolution $R\sim0.003\%$ ($\sigma_{\sqrt s} \sim 2$~MeV) is employed, then the necessary luminosity per scan point is 0.0015~fb$^{-1}$ for $m_h \lsim 2M_W$ and $m_h$ not near $M_Z$. As an example, suppose that the LHC has measured $m_h = 110\rm~GeV\pm100~MeV$. The number of scan points to cover a 200~MeV region in $\sqrt s$ at the FMC is $\rm200~MeV/2~MeV
\sim 100$, and a total luminosity of $100\times (0.0015\rm~fb^{-1}/point) = 0.15~fb^{-1}$ is needed to discover the Higgs and reach an accuracy on its mass of
\begin{equation}
\Delta m_h \simeq \sigma_{\sqrt s} \sim 2\rm\ GeV \,.
\end{equation}

Once $m_h$ is determined to an accuracy $\Delta m_h \sim {\cal O} \left(\sigma_{\sqrt s}\right)$ a three point fine scan can be made with one setting at the apparent peak and two settings on the wings at $\pm\sigma_{\sqrt s}$ from the peak. The ratios of $\sigma({\rm wing}^i)/\sigma({\rm peak}^i)$  determine $m_h$ and $\Gamma_h$. For example, for $m_{h_{\rm SM}}=110$~GeV, $\Gamma_{h_{\rm SM}}=3$~MeV, a 5\% accuracy on $\Gamma_h$
could be achieved with $R=0.003\%$ and $L_{\rm total} = 2\rm~fb^{-1}$.

The heavier neutral MSSM Higgs bosons are also observable in the $s$-channel. Figure~5 give the cross sections and significance of the CP-odd state $A^0$ versus the $A^0$ mass, assuming $R=0.1\%$ and $L=0.01\rm~fb^{-1}$. Discovery and study of the $A^0$ is possible at all $m_A$ if $\tan\beta>2$ and at $m_A<2m_t$ if $\tan\beta\lsim2$. 

\begin{figure}[h]
\centering
\leavevmode
\epsfxsize=3.1in\epsffile{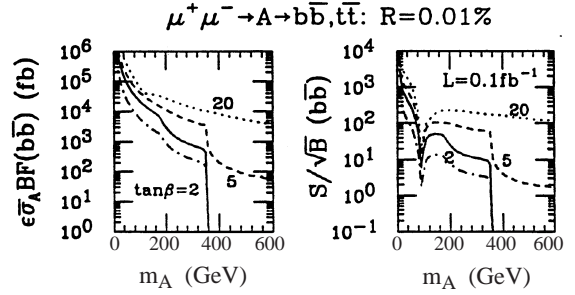}

\caption[]{Cross sections and significance for detection of the $A$ Higgs boson with an efficiency $\epsilon=0.5$ and a luminosity $L=0.1\rm~fb^{-1}$. From Ref.~\cite{prl}.}
\end{figure}

The possibility that $A^0$ and $H^0$ may be nearly mass degenerate is of particular interest for $s$-channel Higgs studies. In the large $m_A$ limit, typical of many supergravity models, the masses of $A^0$, $H^0$ and $H^\pm$ are similar and $h^0$ is similar to $h_{\rm SM}$ in its properties. In this situation the $A^0$ and $H^0$ contributions can be separated by an $s$-channel scan; see Fig.~6.

\begin{figure}[h]
\centering
\leavevmode
\epsfxsize=3.2in\epsffile{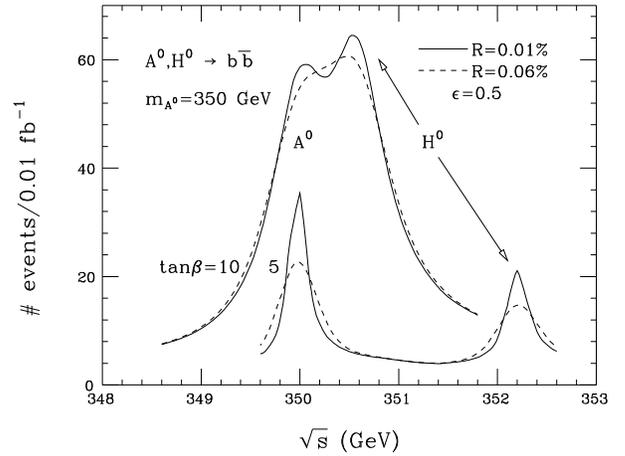}

\caption[]{Separation of $A^0$ and $H^0$ Higgs signals for two values of $\tan\beta$. From Ref.~\cite{physrep}.}
\end{figure}

\section{Threshold Physics at the FMC}

Precision measurements of the $W$-boson and top quark masses provide important tests of radiative corrections through the relation
\begin{equation}
M_W  = M_Z \left[ 1 -{\pi\alpha\over\sqrt 2 \, G_\mu M_W^2 (1-\delta r)} \right]^{1/2} \,,
\end{equation}
where $\delta r$ represents loop contributions\cite{langacker}. In the SM, $\delta r$ depends on $m_t^2$ and $\log m_h$; in the MSSM the sparticle masses also enter in $\delta r$. The optimal relative precision for tests of this relation is $\Delta M_W\approx {1\over140}\Delta m_t$ (for example, $\Delta M_W\approx 6$~MeV and $\Delta m_t\approx 800$~MeV). Precision $M_W$ and $m_t$ measurements can be made at the FMC. With 100~fb$^{-1}$ of luminosity devoted to a measurement of $\sigma(\mu^+\mu^-\to W^+W^-)$ at $\sqrt s = 161$~GeV, an accuracy\cite{mwmt}
\begin{equation}
\Delta M_W= 6\rm\ MeV
\end{equation}
could be realized. Figure~7 shows the $WW$ cross section rise in the threshold region. With 100~fb$^{-1}$ of luminosity to make a 10 point threshold region measurement of $\sigma(\mu^+\mu^-\to t\bar t)$, an accuracy
\begin{equation}
\Delta m_t = 70\rm\ MeV
\end{equation}
could be obtained\cite{mwmt}. Figure~8 shows the $t\bar t$ threshold cross section. The shape of the $t\bar t$ cross section rise with $\sqrt s$ also constrains $\alpha_s$ and the top quark decay width. 
For the above precisions on $M_W$ and $m_t$, radiative corrections would constrain $m_h$ to
\begin{equation}
\Delta m_h =0.12m_h \,,
\end{equation}
which is illustrated in Fig.~9.

\begin{figure}[t]
\centering
\leavevmode
\epsfxsize=3in\epsffile{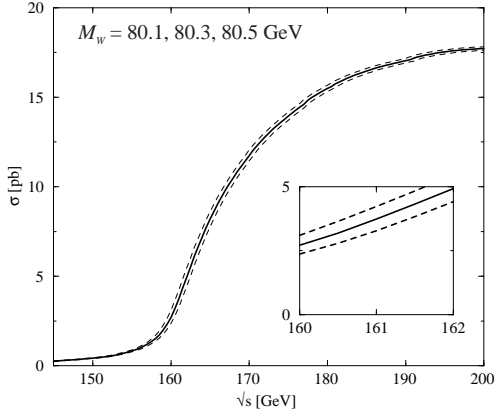}

\caption[]{The $\mu^+\mu^-\to W^+W^-$ cross section in the threshold region for $M_W=80.3$~GeV (solid) and $M_W=80.1$, 80.5~GeV (dashed). From Ref.~\cite{mwmt}.}
\end{figure}

\begin{figure}[h]
\vskip.5in
\centering
\leavevmode
\epsfxsize=3in\epsffile{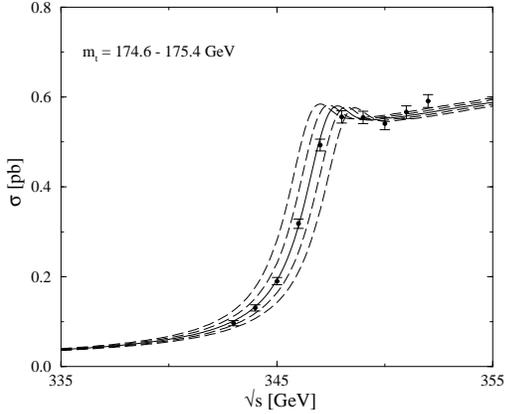}

\caption[]{Threshold cross section for $t\bar t$ production; the curves correspond to shifts of $m_t$ in 200~MeV increments. Sample data are based on a scan with 10~fb$^{-1}$ luminosity at each point. From Ref.~\cite{mwmt}.}
\end{figure}

\begin{figure}[t]
\centering
\leavevmode
\epsfxsize=2.6in\epsffile{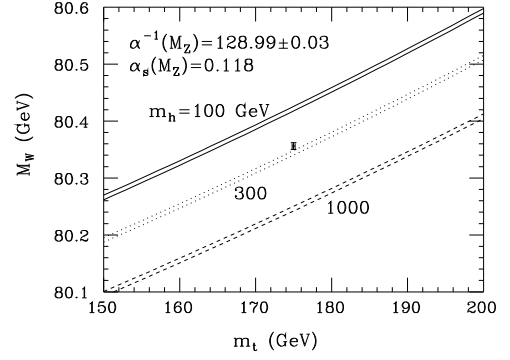}

\caption[]{Correlation of $M_W$ and $m_t$ in the SM with electroweak and QCD corrections. The data point illustrates the potential accuracy of an indirect $m_h$ determination with $M_W=80.356\pm0.006$~GeV and $m_t=175\pm0.2$~GeV. From Ref.~\cite{mwmt}.}
\end{figure}

Measurements of the $\mu^+\mu^-\to Zh$ cross section just above threshold will allow a direct precision measurement of the Higgs mass. The threshold behavior is $S$ wave, so the rise in the cross section is rapid, as shown in Fig.~10. The precision attainable from a 100~fb$^{-1}$ measurement of the $Zb\bar b$ cross section at $\sqrt s = M_Z+m_h + 0.5$~GeV is shown in Fig.~11. For $m_h\sim100~$GeV, the estimated precision is\cite{zh}
\begin{equation}
\Delta m_h \sim 45\rm\ MeV \,.
\end{equation}

\begin{figure}[h]
\centering
\leavevmode
\epsfxsize=3in\epsffile{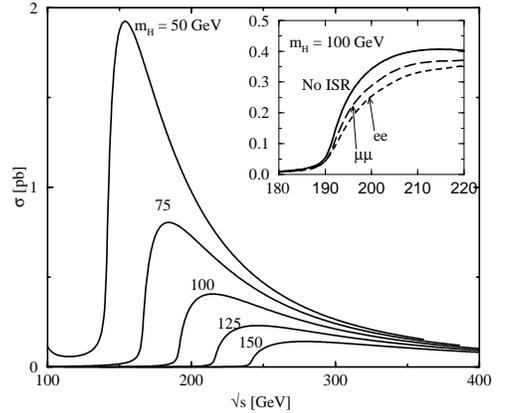}

\caption[]{The $\mu^+\mu^-\to Z^*H\to f\bar fH$ cross section versus $\sqrt s$ for various Higgs boson masses. The insert shows the threshold region for $m_H=100$~GeV. From Ref.~\cite{zh}.}
\end{figure}

\begin{figure}[h]
\centering
\leavevmode
\epsfxsize=3in\epsffile{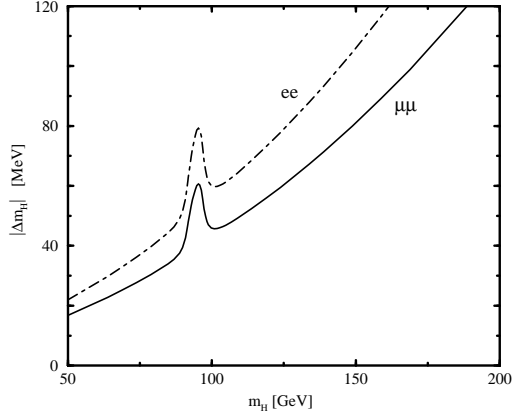}

\caption[]{The precision $\Delta m_H$ attainable from a 50~fb$^{-}$ measurement of the $Zb\bar b$ cross section at $\sqrt s = M_Z+m_H+0.5$ versus $m_H$. From Ref.~\cite{zh}.}
\end{figure}

\section{The High Energy Frontier}

There are compelling motivations from both the weakly and strongly interacting scenarios for constructing a 3--4~TeV $\mu^+\mu^-$ machine.

\subsection{Weak symmetry breaking}

Supersymmetry has many scalar particles (sleptons, squarks, Higgs bosons). Some or possibly many of these scalars may have TeV-scale masses\cite{nelson}. Since spin-0 pair production is $p$-wave suppressed at lepton colliders, energies well above the thresholds are required to have sufficient production rates; see Fig.~12. Moreover, the excellent initial state energy resolution of a muon collier is highly advantageous in reconstructing sparticle mass spectra from their complex decays.

\begin{figure}[t]
\centering
\leavevmode
\epsfxsize=2.5in\epsffile{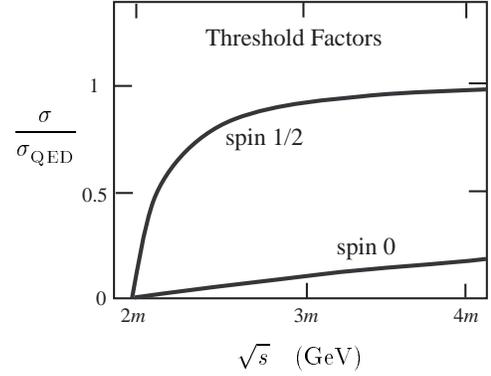}

\caption{Kinematic suppressions of the threshold cross sections of fermion pairs and squark pairs at $e^+e^-$ or $\mu^+\mu^-$ colliders.}
\end{figure}

\subsection{Strong symmetry breaking}

If no Higgs boson exists with $m_h<600$~GeV, then partial wave unitarity of $WW\to WW$ scattering requires that the scattering be strong at the 1--2~TeV energy scale. The $WW\to WW$ scattering amplitude behaves like
\begin{eqnarray}
A &\sim& m_H^2/v^2 \quad\,\ \mbox{if light Higgs} \,,\\
&\sim& s_{WW}/v^2\quad \mbox{if no light Higgs}\,.
\end{eqnarray}
In the latter scenario new physics must be manifest at high energies. Energy reach is a critical matter here with subprocess energies $\sqrt{s_{WW}} \gsim 1.5$~TeV needed to probe strong $WW$ scattering. Since $E_\mu\sim (3\mbox{--}5)E_W$, this condition implies
\begin{equation}
\sqrt{s_{\mu\mu}} \sim (3\mbox{--}5)\sqrt{s_{WW}} \gsim 4\rm\ TeV \,.
\end{equation}
Thus a 4~TeV muon collider would have sufficient energy for a comprehensive study of SEWS.

The nature of the underlying physics will be revealed by the study of all possible vector boson -- vector boson scattering channels, since the sizes of the signals depend on the resonant or nonresonant interactions in the different isospin channels. Several simple models for SEWS are:

\renewcommand{\theenumi}{\alph{enumi}}
\renewcommand{\labelenumi}{\theenumi)}
\begin{enumerate}

\item a heavy scalar particle ($H^0$) with mass $M_S \sim 1$~TeV and width $\Gamma_S \sim 300$--500~GeV,

\item a heavy vector particle ($\rho_{\rm TC}$) with $M_V \sim 1$~TeV and $\Gamma_V \sim 200$~GeV,

\item a nonresonant amplitude $A_{W_LW_L}\sim s/v^2$ (where $v$ is the vacuum expectation value) which is an extrapolation of the low energy theorem (LET). 
\end{enumerate}

\noindent\looseness=-1
The predictions of the three scenarios are distinct:
\begin{equation}
{\sigma(W_LW_L\to Z_LZ_L)\over \sigma(W_LW_L\to W_LW_L)} \quad
\arraycolsep=1em
\begin{array}[b]{rrc}
H^0 & \rho_{\rm TC} & \rm LET\\ \noalign{\vskip1ex}
\displaystyle\simeq{1\over2} & \displaystyle \ll{1\over2} &\displaystyle{3\over2}
\end{array}
\end{equation}
The signals in all the models are impressive in size. For example, in the $H^0$ case,
\begin{eqnarray}
&& \hspace{-.2in} \sigma(W^+W^-\rm\ signal) = 70\rm\ fb\,,\\
&& \hspace{-.2in} \sigma(ZZ\rm\ signal) = 40\rm\ fb\,,
\end{eqnarray}
where
\begin{equation}
\sigma({\rm signals}) = \sigma(m_H = 1{\rm\ TeV}) - \sigma(m_H=0) \,.
\end{equation}
With 1000~fb$^{-1}$ per year the Next Muon Collider will allow comprehensive studies to be made of any SEWS signals. Figure~13 illustrates the $WW$ invariant mass distribution for the models\cite{strongww}. It would be possible to measure the width of a 350~GeV scalar resonance to $\pm30$~GeV and thereby differentiate among scalar models, for example. 

\begin{figure}[t]
\centering
\leavevmode
\epsfxsize=3.2in\epsffile{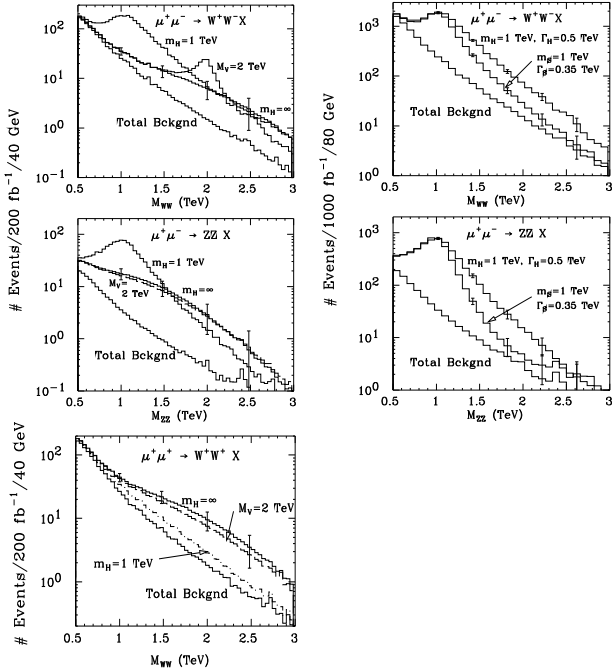}

\caption[]{Events versus the $VV$ invariant mass for two SEWS models (including the backgrounds) and for the backgrounds alone. From Ref.~\cite{strongww}.}
\end{figure}

Angular distributions of the jets in the $WW\to4$\,jet final state will provide a powerful discrimination of SEWS from the light Higgs theory, as illustrated in Fig.~14. Here $\theta^*$ is the angle of the $q$ or $\bar q$ from $W$ decays (in the $W$ rest frame, relative to the $W$ boost direction in the $WW$ c.m.\ frame) averaged over all configurations.

\begin{figure}[t]
\centering
\leavevmode
\epsfxsize=2.5in\epsffile{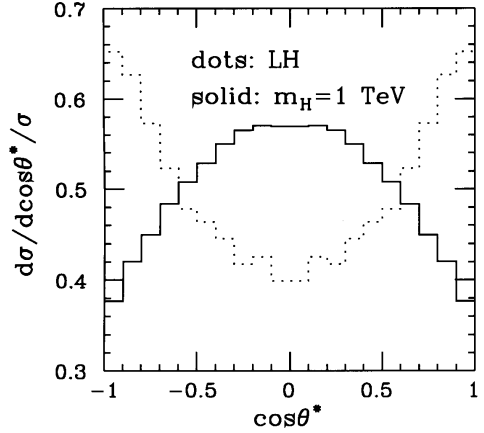}

\caption[]{Normalized event distribution versus $\cos\theta^*$ of the $q$ and $\bar q$ from $W$ decays in the $W^+W^-$ final state. From Ref.~\cite{strongww}.}
\end{figure}

\section{Summary}

In summary, muon colliders offer decisive probes of electroweak symmetry breaking and physics beyond the standard model, including:

\begin{itemize}\addtolength{\itemsep}{-.05in}
\item $s$-channel Higgs production for precision $m_h$ and $\Gamma_h$ measurements,

\item $WW$, $t\bar t$, $Zh$ threshold measurements of $M_W$, $m_t$, $m_h$ to test EW radiative corrections,

\item the discovery and study of heavy MSSM Higgs ($H^-$, $A^0$, $H^\pm$),

\item heavy supersymmetric particle production of $\gsim1$~TeV states $\tilde q$, $\tilde \ell$, $H$, $A$,

\item a strongly interacting $WW$ sector.

\end{itemize}

Muon colliders are smaller and potentially less costly than other colliders of comparable energy\cite{palmer}. They may consequently prove to be an excellent option for exploring the high energy frontier in the 21st century!

\section*{Acknowledgments} 
I thank M.S.~Berger, J.F.~Gunion, and T.~Han for collaborations on the work reported here and T.~Han for helpful discussions during the preparation of this report. This work was supported in part by the U.S.~Department of Energy under Grant No.~DE-FG02-95ER40896 and in part by the University of Wisconsin Research Committee with funds granted by the Wisconsin Alumni Research Foundation.


\begin{thebibliography}{99}

\bibitem{workshops}
{\it Proceedings of the First Workshop on the Physics Potential and Development of $\mu^+\mu^-$ Colliders}, Napa, California, 1992, Nucl.\ Instr.\ and Meth.\ {\bf A350}, 24 (1994); 
{\it Proceedings of the Second Workshop on the Physics Potential and Development of $\mu^+\mu^-$ Colliders}, Sausalito, California, 1994, ed.\ by D.~Cline, American Institute of Physics Conference Proceedings 352 (AIP, New York, 1996);
{\it Proceedings of the 9th Advanced ICFA Beam Dynamics Workshop: Beam Dynamics and Technology Issues for $\mu^+\mu^-$ Colliders}, Montauk, Long Island, 1995, ed.\ by J.C.~Gallardo, American Institute of Physics Conference Proceedings 372 (AIP, New York, 1996);
{\it Proceedings of the Symposium on Physics Potential and Development of $\mu^+\mu^-$ Colliders}, San Francisco, California, 1995, ed.\ by D.~Cline and D.~Sanders, Nucl.\ Phys.~B Proceedings Supplement {\bf 51A} (1996);
{\it $\mu^+\mu^-$ Collider: A Feasibility Study}, 1996 DPF/DPB Summer Study on New Directions for High-Energy Physics, Snowmass, Colorado, June 25--July 12, 1996, Brookhaven National Laboratory publication BNL-52503 (1996);
R.B.~Palmer and J.C.~Gallardo, in {\it Techniques and Concepts of High Energy Physics IX}, ed.\ by T.~Ferbel, Plenum Press (1997) [acc-phys/9702017].



\bibitem{physrep}
V.~Barger, M.S.~Berger, J.F.~Gunion, and T.~Han, Phys.\ Rep.\ {\bf 286}, no.~1, p.~1 (1997).


\bibitem{saus}
 V.~Barger et al., in {\it Proceedings of Second Workshop on Physics Potential and Development of $\mu^+\mu^-$ Colliders},  Sausalito, California, 1994, ed.\ by D.~Cline, American Institute of Physics Conference Proceedings 352, p.~55 (AIP, New York, 1996).

\bibitem{prl}
V.~Barger, M.S.~Berger, J.F.~Gunion, and T.~Han, Phys.\ Rev.\ Lett.\ {\bf 75}, 1462 (1995).

\bibitem{langacker}
See e.g.\ P. Langacker and J.~Erler, in {\it Review of Particle Physics}, Phys.\ Rev.\ {\bf D54}, 85 (1996); K.~Hagiwara, D.~Haidt, and S.~Matsumoto, hep-ph/9706331 (1997).

\bibitem{mwmt}
V. Barger, M.S. Berger, J.F. Gunion, and T. Han, Phys.\ Rev.~{\bf D56}, 1714 (1997).

\bibitem{zh}
V. Barger, M.S.~Berger, J.F.~Gunion, and T.~Han, Phys.\ Rev.\ Lett. {\bf 78}, 3991 (1997).

\bibitem{nelson} See e.g.\ S.~Ambrosiano and A.~Nelson, University of Washington report UW-PT-97-11 [hep-ph/9707242] (1997).

\bibitem{strongww}
V. Barger, M.S. Berger, J.F. Gunion, and T. Han, Phys.\ Rev.\ {\bf D55}, 142 (1997).

\bibitem{palmer}
See e.g.\ R.B.~Palmer and J.C.~Gallardo, Brook\-haven National Laboratory report BNL-64148 (1997).

\end{thebibliography}
\end{document}